# Emergence of Active Inference from a Chemical Oscillator: A Constructive Approach to Pre-genetic Homeostasis

Takeshi Ishida[1]*

[1]Department of Ocean Mechanical Engineering, National Fisheries University; Shimonoseki, 759-6595, Japan.

*Corresponding author. Email: ishida@fish-u.ac.jp

**Abstract:** How could primordial life, before the evolution of genetic systems, adapt to fluctuating environments and achieve homeostasis? This study proposes a minimal, chemically plausible model where homeostasis emerges from a simple chemical reaction network. It utilizes an internal Lotka-Volterra chemical oscillator as a "search engine" to periodically vary a protocell's pigmentation. The system then optimizes its internal state by evaluating the temporal correlation between these internal fluctuations and a single global metric—the cell's self-replication rate—through a mechanism termed "antagonistic memory molecules." Numerical simulations demonstrate that the model can autonomously converge to and maintain the optimal temperature for its self-replication, even amidst significant environmental fluctuations. These findings provide a constructive proof-of-concept for how a core process of active inference can emerge from a simple physicochemical system, offering a concrete scenario for the acquisition of adaptive capabilities at the origin of life.

**Background**
*Pre-genetic Adaptation: A Fundamental Question in the Origin of Life*
   The capacity to maintain internal homeostasis and persist within a constantly changing environment is a fundamental characteristic of life. In extant organisms, this capacity is underpinned by an intricate genomic blueprint, realized through mechanisms such as feedback control of gene expression and epigenetic modifications (1, 2). However, given that this genetic system is itself a product of Darwinian evolution over vast timescales, a critical question emerges: How did primordial life, prior to the establishment of a stable genetic blueprint, adapt to its environment and acquire homeostasis?
   Darwinian evolution is a powerful optimization process, but it operates on a relatively slow timescale, driven by random mutations and subsequent natural selection. Consequently, this process is too slow to allow an individual organism to respond to rapid environmental changes within its lifetime. This suggests that a more rapid adaptive mechanism, one not directly reliant on genes, must have existed from the earliest stages of life. This notion is supported by the work of Kussell et al. (3), who demonstrated that the ability to flexibly alter phenotypes is critical for survival in fluctuating environments. Exploring the physicochemical basis of this "rapid, non-genetic adaptation" is therefore an indispensable step toward understanding the origin of life.

*Theoretical Framework: The Free Energy Principle and the Bottom-Up Challenge*
   In recent years, the Free Energy Principle (FEP), proposed by Karl Friston, has gained prominence as a unifying framework for understanding self-maintenance and adaptation in living systems (4). According to the FEP, living systems must maintain their existence by acting to minimize their long-term "surprise"—or prediction error—relative to their own generative model of the environment. This minimization of prediction error is achieved through two complementary pathways:

- **Perception**: Updating the internal model (or beliefs) to better align with sensory inputs.
- **Action**: Intervening in the world to alter sensory inputs, thereby making them conform to the system's predictions.

This process of inference through action is known as "Active Inference" (4).

The scope of the FEP extends beyond the cognitive functions of the brain. Recent theoretical advances have expanded its application, offering new perspectives on the behavior of biological systems that lack a nervous system. For instance, biological phenomena such as morphogenesis in multicellular organisms (5) and chemotaxis in single-celled organisms (6, 7) are being reinterpreted through the lens of active inference.

This expansion of the FEP resonates deeply with the burgeoning field of "Basal Cognition" (8). Basal cognition challenges the traditional neurocentric view that cognitive functions such as thinking, learning, and decision-making are exclusive to organisms with brains. Instead, it explores the remarkable problem-solving abilities demonstrated by non-neuronal life forms such as bacteria, slime molds, plants, and the somatic cell colonies that constitute animal bodies (9). This approach posits that cognition is not a later evolutionary addition but a fundamental property inherent to life itself.

The FEP serves as a normative (top-down) theory for the phenomena uncovered by basal cognition, providing a unified mathematical and physical foundation to explain *why* and *how* all living systems behave as they do (7). However, while the FEP specifies the principles that living systems ought to follow, it leaves a significant bottom-up question unanswered: "How does such teleological (goal-directed) behavior—that is, basal cognition—self-organize and 'emerge' from simple physicochemical systems?" This remains a major mystery. Bridging this gap between the normative theory and physicochemical reality is crucial for a deeper understanding of both the origin of life and the principles of intelligence.

***Our Approach: Emergence of Active Inference through Chemical Computing***

To address the challenge outlined above, this study adopts an approach from the framework of "chemical computing." Chemical computing, a field that utilizes molecular interactions for information processing, is broadly divided into two major paradigms. The first is the "Design" approach, which aims to meticulously engineer molecular-level equivalents of digital computer logic. The second is the "Emergence" approach, which instead seeks to leverage the intrinsic properties of chemical reaction systems, such as nonlinearity and self-organization, to enable the autonomous emergence of adaptive behaviors (10, 11).

This research is positioned within the latter "Emergence" paradigm. Rather than implementing specific logic gates, this study explores how a homeostatic function equivalent to active inference can self-organize from a simple chemical reaction network. To this end, we propose a constructive model, the "Minimal Cognition Engine," to test the following hypothesis: a system can autonomously discover and maintain an optimal state without explicit goals or direct sensory information by utilizing a simple chemical oscillator as a "search engine" to actively explore its parameter space and adjusting its internal state based solely on the temporal correlation between these explorations and the resulting overall system success (e.g., self-replication rate).

In this paper, we demonstrate through simulation a concrete mechanism by which a pre-genetic, primordial life form could achieve homeostasis in a fluctuating environment via the simple loop of "exploration through oscillation and optimization through correlation". We thereby show that this model serves as a minimal, chemically plausible, constructive model of basal cognition that embodies the FEP, offering new insights into its potential physicochemical underpinnings.

**Overview of the Minimal Cognition Engine**

This study proposes a constructive model to investigate the minimal mechanisms by which a primordial chemical system, devoid of genetic information and complex sensory organs, could adapt to its environment and acquire homeostasis. The model comprises two core components: a "body" that physically interacts with the environment, and an "adaptive mechanism" that controls the body.

*Physical Model: A Single-Celled Organism in a Daisyworld-like Environment*

The physical environment of the model is a hypothetical planet inspired by Lovelock's Daisyworld model (12).
- Environment: The planet orbits a sun with periodically fluctuating luminosity and features oceans on its surface. The ocean water temperature varies in response to fluctuations in solar luminosity.
- Agent: The ocean surface is populated by single-celled organisms (agents) devoid of genes. Each agent can continuously vary its surface pigmentation, from white to black, based on its internal chemical state. This level of pigmentation determines the cell's albedo (solar reflectivity) and, consequently, its own temperature.
- Survival and Replication: The cell's self-replication rate depends on its temperature. The rate is maximized at a specific optimal replication temperature, $T_{opt}$, and decreases at higher or lower temperatures, following a concave-down parabolic curve. The cells proliferate according to this replication rate while also dying at a constant, temperature-independent rate.

The model's objective is to demonstrate the emergence of a homeostatic function that allows the agents to autonomously maintain the optimal temperature, $T_{opt}$, which maximizes their replication rate. This must be achieved without the agents directly sensing either their own cellular temperature or the external environmental temperature.

*Adaptive Mechanism: The Antagonistic Memory Molecule Model*

To achieve homeostasis, the agents employ an adaptive mechanism named the "Antagonistic Memory Molecule Model". This model was developed to achieve self-optimization through a continuous, chemically plausible process, moving beyond an earlier algorithm-based approach that relied on conditional logic (see Appendix A-2). By minimizing such algorithmic branching, the model achieves self-optimization through continuous dynamics.

This mechanism consists of three core components (see Fig. 1):

**1. Chemical Oscillator as a Search Engine:** The model assumes an internal chemical reaction network within the cell, described by the Lotka-Volterra (LV) equations. This network functions as a chemical oscillator with periodically varying concentrations of chemical species *X* and *Y*. The LV system was chosen for its simplicity and its capacity for the oscillation center to be shifted by altering parameters such as reaction rates. By utilizing this ability, the oscillator functions as a "search engine" to actively explore the system's parameter space. Specifically, the production rate of an oscillator product, *P*, drives cyclic variations in the cell's pigmentation (albedo). (For details, see Appendix A-1).

**2. Physical Interaction:** The output of the oscillator (the production rate of $P$) alters the cell's physical property of pigmentation (albedo). This allows the cell to actively vary its own temperature, which in turn affects its self-replication rate. Through this process, the internal chemical "exploration" generates a physical "action" and a corresponding survival-relevant "outcome." (For details, see Appendix A-2).

**3. Memory Molecules for Correlation-Based Learning:** At the core of the adaptation are two types of hypothetical "memory molecules," $M_{good}$ and $M_{bad}$. These molecules chemically accumulate evidence of whether the cell's state is trending in a "good" or "bad" direction.

- *Production of Memory Molecules:* The production of these molecules is controlled by the temporal correlation between the trend in the internal state (change in $P$ production rate) and the trend in the resulting success (change in self-replication rate, $\beta$). When the directions of change align (e.g., an increase in $P$ production leads to an increase in $\beta$), reactions that produce $M_{good}$ are promoted. Conversely, when the directions are opposed, reactions producing $M_{bad}$ are promoted.
- *Parameter Adjustment:* The concentration balance between $M_{good}$ and $M_{bad}$ continuously adjusts the LV oscillator's parameters (reaction constant $k_1$) and the baseline pigmentation (albedo center value). For instance, if the concentration of $M_{good}$ exceeds that of $M_{bad}$, the current direction of exploration is deemed "correct," and the parameters are updated to reinforce exploration in that direction.

This continuous loop of "exploration through oscillation" and "correlation-based learning via memory molecules" chemically implements the essential process of active inference—learning from past experience to optimize future actions—without requiring rigorous probabilistic calculations. (For details, see Appendix A-3).

*Chemical Plausibility*

The proposed "Antagonistic Memory Molecule Model" was designed for both simplicity and chemical feasibility. The model's correlation-based learning mechanism is situated within a broader line of research, including the work by McGregor et al. (13) on the emergence of associative learning in chemical reaction networks, which demonstrates that simple chemical systems can acquire the ability to modify their behavior based on experience.

Two key aspects underscore the model's plausibility:

- **Biochemical Analogy:** The process by which the memory molecules $M_{good}$ and $M_{bad}$ adjust the parameter $k_1$ is analogous to feedback mechanisms universally found in biochemical reactions, such as allosteric regulation (14) and enzyme activation/inactivation (15).
- **Description by the Law of Mass Action:** The entire model can be described by the law of mass action as a system of coupled differential equations governing the production and degradation of each chemical species (for details, see Appendix A-3). This formulation suggests that the adaptive mechanism presented in this model can emerge from the fundamental principles of chemical kinetics without requiring complex decision-making logic.

This paper will demonstrate through simulation that this chemically plausible model enables the emergence of homeostasis, allowing for the autonomous maintenance of an optimal temperature under a fluctuating external environment.

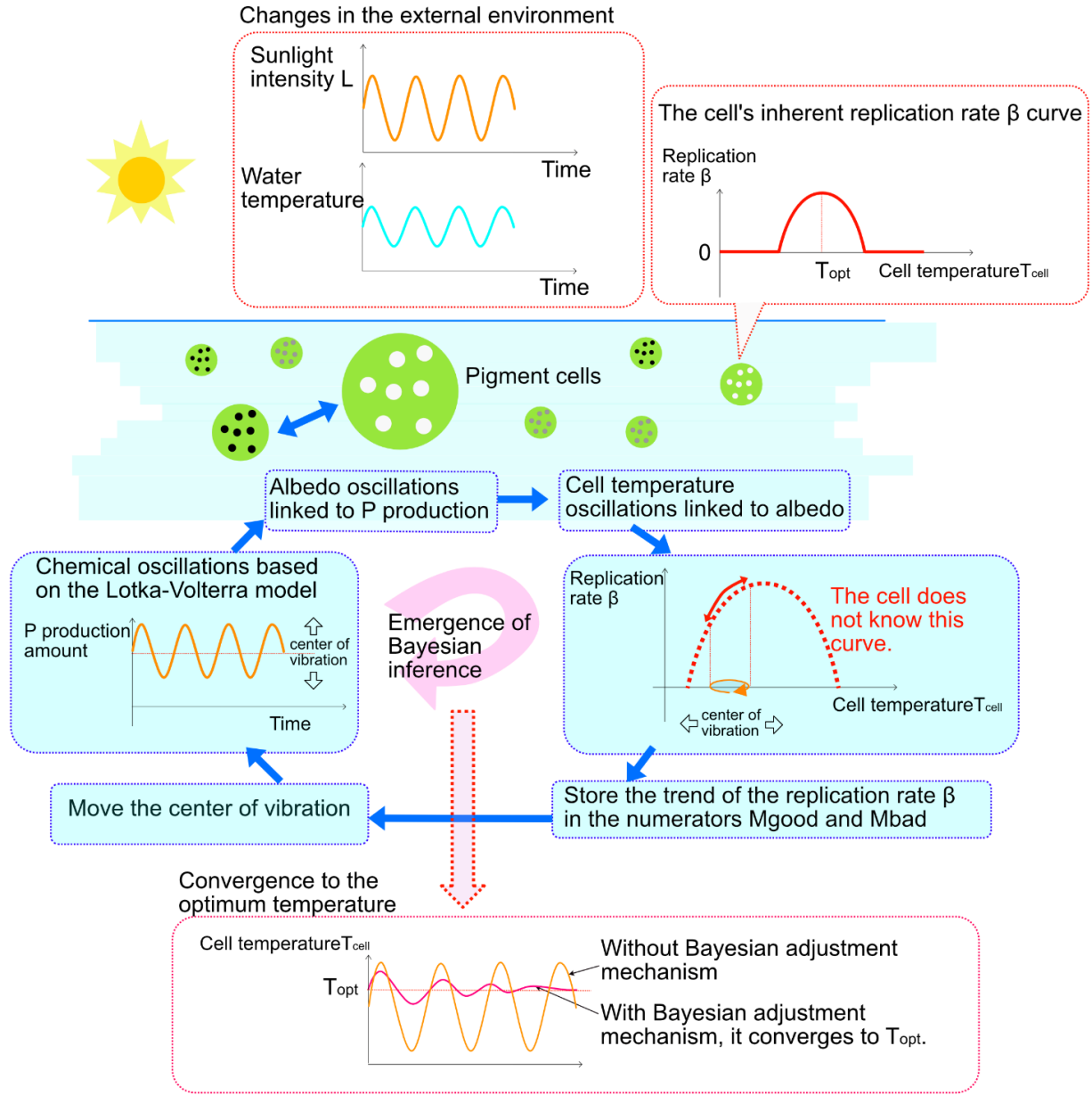

**Fig. 1. Schematic overview of the model.** The model features a population of single-celled, non-genetic agents on the surface of an ocean on a hypothetical planet. The environment is characterized by periodically fluctuating solar luminosity and water temperature. The self-replication rate ($β$) of each cell is temperature-dependent, maximized at an optimal temperature ($T_{opt}$) and decreasing otherwise. Internally, a Lotka-Volterra (LV) chemical oscillator drives cyclic changes in pigmentation (albedo) by altering the production rate of a substance $P$. This allows the cell to actively modulate its own temperature, which in turn affects its replication rate. This entire process forms a feedback loop where internal exploration (oscillation) generates an action (pigmentation change) that influences a success metric (replication rate), thereby enabling the system to converge toward the optimal temperature.

**Results**

This section presents the results of simulations conducted to test whether the "Antagonistic Memory Molecule Model," as proposed in previous section, can lead to the emergence of homeostatic functions. To objectively evaluate the model's performance, we define two quantitative metrics, calculated after the simulation reaches a steady state:

- **Accuracy:** Defined as the time-averaged value of the cell temperature ($T_{cell}$) in the steady state. This metric assesses the proximity of the cell temperature to the target optimal replication temperature ($T_{opt}$).
- **Stability:** Defined as the standard deviation of the cell temperature ($T_{cell}$) in the steady state. This metric quantifies the effectiveness of the homeostatic mechanism in suppressing external environmental fluctuations and maintaining a stable internal state.

Using these metrics, we tested the following three hypotheses.

*Hypothesis 1: Emergence of a Basic Homeostatic Function*

[Hypothesis 1] The model can generate a basic homeostatic function by autonomously converging to the optimal replication temperature ($T_{opt}$), which maximizes its replication rate, even under a fluctuating external environment.

To test this fundamental hypothesis, a simulation was performed using a standard set of parameters: an optimal replication temperature of $T_{opt}$ = 22.0 °C, a mean sunlight intensity of 0.7, and an amplitude of 0.21. Fig. 2A shows a representative result of this simulation. After the simulation begins, the cell temperature ($T_{cell}$, orange line) clearly converges toward the optimal replication temperature ($T_{opt}$ = 22.0 °C) through the adaptive learning mechanism, despite the influence of the periodically fluctuating external water temperature ($T_{water}$, light blue dashed line). This convergence results from the automatic adjustment of the adaptive parameters—the central albedo value and the number of memory molecules—to values that maximize the self-replication rate (see Appendix A-5-1 for details).

Particularly during the initial phase of the simulation (steps 0–200), the parameters fluctuate significantly as they search for the optimal value. This behavior is attributable to the adjustment from the initial values of the memory molecules. Subsequently, the system transitions to a stable state that maintains a temperature around $T_{opt}$. This result strongly supports Hypothesis 1, demonstrating that the model can achieve homeostasis based on the simple principle of "exploration by oscillation and optimization by correlation" without being explicitly given a target value.

*Hypothesis 2: Adaptability to Diverse Target Temperatures*

[Hypothesis 2] The model's homeostatic function is not fixed to a specific target; it can adapt flexibly to different optimal replication temperatures ($T_{opt}$).

To test the model's adaptive flexibility, the optimal replication temperature ($T_{opt}$) was varied from 20.0°C to 24.0°C. The results, presented in Fig.2B, show that the cell temperature successfully adjusts to the corresponding optimal temperature in each case.

Table 1 summarizes the mean (Accuracy) and standard deviation (Stability) of the cell temperature for each $T_{opt}$ setting, calculated during the final phase of the simulation (from time step 801 to 1000). In all settings, the mean cell temperature closely approximated the target $T_{opt}$, demonstrating the model's high accuracy. Furthermore, the standard deviation remained stable across all cases, converging to approximately 0.6°C.

These results support Hypothesis 2, demonstrating the model's capability to self-optimize for diverse target states without high dependence on specific parameter values.

A Emergence of a basic homeostatic function

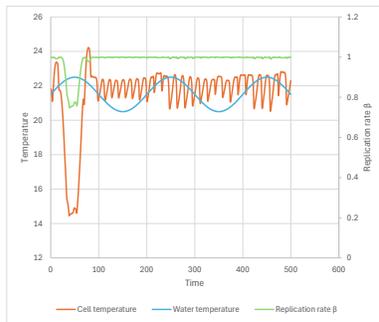

B Adaptation to diverse optimal replication temperatures ($T_{opt}$)

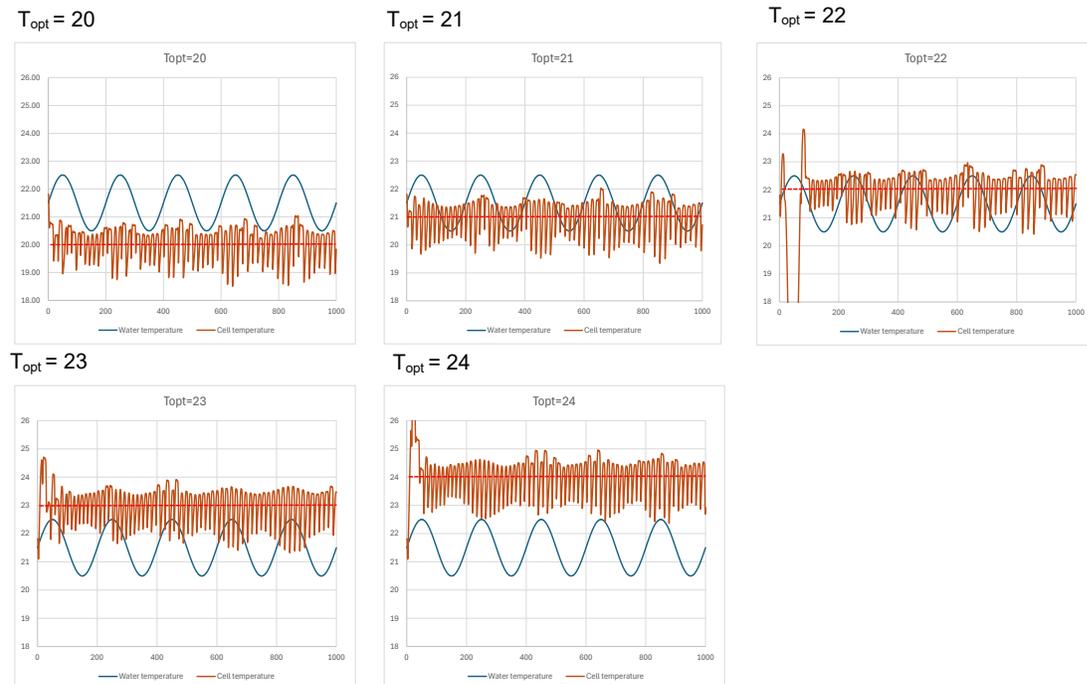

**Fig. 2. Emergence of homeostatic functions. (A) Emergence of a basic homeostatic function.** The cell temperature ($T_{cell}$, orange line) converges toward the optimal replication temperature ($T_{opt}$ = 22.0 °C) despite periodic fluctuations in the external water temperature ($T_{water}$, light blue dashed line). This convergence is achieved through the automatic adjustment of adaptive parameters, such as the central albedo value and memory molecule concentrations, to values that maximize the self-replication rate. **(B) Adaptation to diverse optimal replication temperatures ($T_{opt}$).** The figure shows simulation results for cases where $T_{opt}$ was varied from 20.0°C to 24.0°C, demonstrating that the cell temperature adjusts to track the respective target in each case.

**Table 1. Mean and standard deviation of cell temperature for various optimal set temperatures ($T_{opt}$).**

| Optimal Temperature | Average cell temperature | Standard deviation of cell temperature |
|---|---|---|
| 20 | 20.08 | 0.607 |
| 21 | 21.05 | 0.620 |
| 22 | 22.06 | 0.597 |
| 23 | 22.99 | 0.633 |
| 24 | 24.00 | 0.617 |

*Hypothesis 3: Robustness Against Environmental Fluctuations*
[Hypothesis 3] The model's homeostatic function remains robust, not failing even under more severe environmental fluctuations.

Finally, we tested the ability of the model's homeostatic function to operate under more demanding conditions. In this test, the system's response was evaluated while varying the amplitude of sunlight intensity, a key factor in environmental fluctuation. As shown in Table 2, the model demonstrates high robustness. Even as the amplitude of sunlight fluctuations was increased from 0.1 to 0.4, the stability (standard deviation) of the cell temperature remained low. Importantly, even under the most severe condition tested (a 0.4 amplitude, corresponding to a water temperature fluctuation of approximately 5°C), the homeostatic function was not compromised, and the cell temperature was maintained around the optimal temperature.

These results strongly support Hypothesis 3, demonstrating that the model is not only functional but is also robust enough to maintain its adaptive capability and preserve homeostasis against significant changes in the external environment. (For the influence of other parameters, see Appendix A-5-2).

**Table 2. Mean and standard deviation (°C) of cell temperature for varying amplitudes of sunlight fluctuation.**

| Solar brightness amplitude | Optimal Temperature | Average cell temperature | Standard deviation of cell temperature |
|---|---|---|---|
| 0.1 | 22 | 22.18 | 0.568 |
| 0.2 | 22 | 22.06 | 0.597 |
| 0.3 | 22 | 22.35 | 0.544 |
| 0.4 | 22 | 22.26 | 0.595 |

**Discussion**

This study has demonstrated through simulation that a simple chemical reaction network can lead to the emergence of homeostasis in a fluctuating environment, based solely on the principle of "exploration by oscillation and optimization by correlation," without reliance on externally provided genetic information or predefined target values. These findings extend beyond the mere presentation of a hypothetical model of life; through the underlying adaptive mechanism, they offer significant implications for multiple disciplines, including theoretical biology, the origin of life, and information science. In this section, we discuss the significance of our findings from four distinct perspectives.

*Biological Plausibility of the Adaptive Mechanism: An Analogy with Bacterial Chemotaxis*

The adaptive mechanism in this model is a remarkably simple heuristic that relies solely on the correlation between two time-series trends: the trend of the internal state and the resulting trend in the system's success (the self-replication rate). This algorithm is consistent with adaptive strategies observed in living organisms.

The most prominent example is the chemotaxis of *Escherichia coli*. *E. coli* does not calculate its optimal direction by spatially differentiating a nutrient gradient. Instead, it employs a search-biasing strategy based on temporal correlation. Following its own movement (a "run" or a "tumble"), the bacterium evaluates only temporal changes—in essence, asking whether the situation has improved recently. If conditions are improving (i.e., the nutrient concentration is increasing), it increases the probability of continuing its straight-line movement (a "run").

The accumulation of the "memory molecules," $M_{good}$ and $M_{bad}$, in our model can be interpreted as a chemical-level implementation of this bacterial strategy. This strong analogy with a well-established biological algorithm suggests that the model is not merely a thought experiment but represents a realistic and powerful adaptive strategy that could have been employed by primordial life before the establishment of genetic information.

*The Chemical Oscillator as a Search Engine: An Analogy with Robotic CPGs*

This research offers a new perspective on the functional role of chemical oscillators. In previous studies, such as those using the Belousov-Zhabotinsky (BZ) reaction, the propagation of chemical waves was often used to perform computations or implement logic gates (17). In contrast, the Lotka-Volterra (LV) oscillator in this model functions not as the final computational output, but as a "search engine" that drives adaptive learning. This autonomous and continuous internal fluctuation actively generates the "action" and "outcome" data that the learning mechanism evaluates.

This concept is strongly analogous to the Central Pattern Generator (CPG) in robotics (18). Just as a CPG generates rhythmic motor patterns for animal locomotion, the LV oscillator in our model generates a rhythmic "metabolic search." However, a key difference exists: while adaptive CPGs require direct sensory feedback, such as joint angles or ground contact, the agent in our model learns by relying solely on a single, global success metric—its self-replication rate. This remarkable simplicity of information input significantly enhances the plausibility of this mechanism functioning within a primordial chemical system possessing only limited sensory capabilities.

*As a Model for the Chemical Emergence of Active Inference*

This model can be interpreted as a concrete "bottom-up" or "constructive" model of the Free Energy Principle (FEP), proposed by Karl Friston, and particularly of active inference. The model's behavior can be viewed as a process of minimizing "prediction error" (the imbalance between $M_{good}$ and $M_{bad}$) in its sensory input (the replication rate, $\beta$) by changing its "action" (the center of albedo oscillation), while simultaneously updating its "internal model" (parameters such as $k_1$ and pigment levels). This constitutes the core loop of "minimizing prediction error through action," a process fundamental to active inference. The correspondence between the components of our model and the key concepts of the FEP is detailed in Appendix A-6.

Many applied FEP studies take a top-down approach, presupposing goals (technically, prior distributions) that are pre-encoded in neural circuits or genes. In contrast, our model starts from simple chemical and physical rules without explicit goals or prior distributions, demonstrating that FEP-like behavior (i.e., the minimization of variational free energy, leading to homeostasis) *emerges* as a result. This presents a scenario that addresses a more fundamental question of great importance in the FEP context: the "origin of priors." That is, how does a living system discover and learn its own preferred states (setpoints)? This interpretation is not merely an analogy. As shown by Poole et al. (19), a chemical reaction network following the law of mass action can itself function as a Bayesian inference machine that autonomously learns a generative model of the world. Our Antagonistic Memory Molecule model can be interpreted as a physical implementation of this abstract theory, applied to the concrete problem of biological homeostasis.

Furthermore, the model suggests that this learning algorithm can be implemented as a concrete chemical reaction network governed by the law of mass action. As recent research indicates, chemical kinetics can itself be interpreted as a form of Bayesian inference (19), and our model attempts to build a concrete bridge between this abstract theory and physical reality. The essence of the discrete-time decision logic implemented in our simulation (i.e., generating $M_{good}$ if the trends of change align) can be captured by the following continuous differential equations:

$$\frac{d[M_{good}]}{dt} = k_p \cdot \sigma(c\dot{x}\dot{y}) - k_d \cdot [M_{good}]$$
$$\frac{d[M_{bad}]}{dt} = k_p \cdot \sigma(-c\dot{x}\dot{y}) - k_d \cdot [M_{bad}]$$

Here, $[M_{good}]$ and $[M_{bad}]$ are the concentrations of the memory molecules, $x$ is the rate of change of the internal state ($P$ production rate), and $y$ is the rate of change of the success metric (replication rate $\beta$). The term $\sigma(z)$ is a sigmoid function that provides an abstract representation of the algorithm and serves as a continuous approximation of the conditional logic used in the simulation (i.e., generating $M_{good}$ if the directions of the trends match). The coefficient $c$ determines the steepness of the switch. The negative sign in the $M_{bad}$ production term, $\sigma(-cxy)$, ensures that this term becomes large when the trends are opposing. This mechanism can be implemented concretely by a common biochemical motif: multi-input control by an allosteric enzyme, an example of which is shown in Appendix A-7.

On the other hand, the search mechanism proposed in this study has limitations inherent in its simplicity. The Lotka-Volterra oscillator used as the search engine is, in essence, a deterministic periodic oscillator. This functions efficiently under simple conditions where the fitness landscape is unimodal—that is, where only a single optimal solution exists. However, in more realistic biological scenarios with multiple favorable states (local fitness maxima), such as different nutrient sources or temperature conditions, this deterministic search is highly likely to become trapped in the local optimum closest to its starting point. This represents a significant limitation to the model's generality, and a mechanism for escaping such local optima is indispensable for considering the emergence of higher cognitive functions. The key to overcoming this problem is the introduction of stochastic fluctuations, as we will discuss in next section.

### *A Model Bridging Active Inference and Basal Cognition*

The "Minimal Cognition Engine" proposed in this study is more than just an optimization algorithm for achieving homeostasis. By reinterpreting its operating principle through the lens of important theoretical frameworks in modern cognitive science, such as the Free Energy Principle (FEP) and basal cognition, a deeper significance emerges, shedding light on the origin of life's intelligence. Our model can be positioned as a constructive model that bridges different hierarchical levels: the universal principles described by the FEP (the "Why"), the biological phenomena revealed by basal cognition (the "What"), and the underlying physicochemical implementation (the "How").

This chemical implementation of active inference embodies, in a remarkably minimal form, the core features of "basal cognition"—a field exploring the learning and decision-making capabilities of non-neuronal organisms. Research in basal cognition has revealed a toolkit of cognitive abilities in these organisms, including goal-directedness, learning, and memory (8,9).

- **Goal-Directedness:** The agent behaves autonomously to maximize an intrinsic success metric—its replication rate, $\beta$—without explicit knowledge of the target value, the optimal temperature $T_{opt}$. This represents the most fundamental form of goal-directedness, where goal-oriented behavior emerges from rules that themselves have no explicit goal.
- **Learning and Memory:** The "antagonistic memory molecules" accumulate the correlation between past actions and their outcomes as chemical concentrations. This represents a simple form of learning that alters behavioral biases through experience, with the concentrations themselves functioning as a short-term chemical memory.

**Embodied Computation:** Cognition (the discovery of the optimal temperature) is a dynamic process emerging from the entire loop of internal chemical reactions, the body (albedo), and the environment, requiring no centralized processor. This adaptive process embodies the concept of "embodied cognition" as proposed by Varela et al. (20).

The preceding discussion offers important implications for the origin and evolution of life. Our model demonstrates how, in a "Metabolism-First" world preceding the establishment of a genetic system, a self-sustaining metabolic cycle (the LV oscillator) could have acquired information-processing and learning capabilities.

Furthermore, the "memory molecules" in our model can be regarded as an analogue of primitive epigenetic memory. These molecules hold information not as a polymer sequence but as a chemical "state," which in turn controls the phenotype (e.g., $k_1$ and albedo). This chemical state can be considered a form of "soft inheritance" that could be transmitted to daughter cells during cell division. It is conceivable that such unstable adaptive marks could, through chemical instability, be converted into and fixed as more stable genetic mutations (21). In this sense, the rapid individual learning demonstrated by our model suggests a potential mechanism capable not only of waiting for Darwinian evolution but also of directing and accelerating it.

**Conclusion and Future Directions**

In this study, we proposed a chemically plausible, constructive model that uses an internal chemical oscillator as a search engine, adjusting its parameters based solely on the temporal correlation between its activity and the overall success of the system. Simulation results demonstrated that this simple mechanism enables the autonomous emergence of homeostasis in a fluctuating environment, without direct knowledge of target values.

These findings not only present a concrete scenario for how primordial life could have adapted to its environment but also serve as a minimal model for how the cognitive loop underlying active inference can emerge from simple physicochemical processes. The contributions of this research extend beyond theoretical biology's exploration of the origin of life and principles of intelligence, offering insights for engineering fields where self-adaptive system design is required.

- **Introducing the Exploration-Exploitation Trade-off:** The current model continues to explore (oscillate) even after reaching an optimal state. A future goal is to chemically implement the "exploration-exploitation trade-off," a universal challenge in adaptive systems. We aim to develop a more sophisticated mechanism by introducing a feedback loop where a "stabilizing molecule S," which suppresses the LV oscillator, is produced when the success rate (replication rate, $β$) is high. This would allow the system to reduce exploration and enhance stability once an optimal state has been found.
- **Expansion into Spatial Dimensions:** While this study used a simplified one-dimensional model for proof-of-principle, we plan to extend this adaptive mechanism to 2D or 3D reaction-diffusion systems or cellular automaton models. This will allow us to investigate how morphogenesis, driven by self-optimization, interacts with homeostatic functions under spatial constraints.
- **Exploring Chemical and Physical Implementation:** The ultimate validation of this model lies in its physical implementation. The logical structure of the "antagonistic memory molecule model" could, in principle, be constructed *in vitro* using, for example, DNA strand displacement (DSD) reactions. Verifying whether the self-optimization phenomenon demonstrated in our simulations can be physically reproduced by constructing such a "wet" artificial life system is a highly challenging yet crucial future task. Furthermore, compartmentalizing this chemical system within protocells and examining its impact on evolution would be a promising line of research for deepening our understanding of the origin of life (22).

**Acknowledgments:** The author wishes to acknowledge the use of the generative AI Gemini for English language proofreading..

**Funding:** This research was supported by grants from Japan Society for the Promotion of Science, KAKENHI Grant Number 19K04896.

**Author contributions:** T.I. conceptualized the study, developed the model, conducted the simulations, analyzed the results, and wrote the manuscript..

**Competing interests:** The author declares no conflict of interest regarding the publication of this paper.

**Data and materials availability:** All data supporting the findings of this study are available within the main text and its supplementary materials.

**Supplementary Materials**

Appendix (Methods, Supplementary Results, and Supplementary Discussion)

# Appendix

## Appendix A-1: The Lotka-Volterra Chemical Oscillation Model
### A-1-1. Overview of the Standard Form of the Lotka-Volterra Equations

The standard form of the Lotka-Volterra equations, which represents the most classic predator-prey model, is a system of nonlinear differential equations describing the change over time in the population (or concentration) of two biological or chemical species.

    $X$: the population or concentration of the prey (or a self-replicating chemical species).

    $Y$: the population or concentration of the predator (or a chemical species that consumes $X$ to replicate).

The equations are expressed as follows:

$$\frac{dX}{dt} = X(a - bY)$$

$$\frac{dY}{dt} = Y(cX - d)$$

Here, the parameters have the following meanings:

    $a$: the intrinsic growth rate of the prey, $X$ (related to its growth rate in the absence of the predator, $Y$; $a>0$).

    $b$: the efficiency of predation of the prey, $X$, by the predator, $Y$ ($b>0$).

    $c$: the efficiency of the increase in the predator, $Y$, from consuming the prey, $X$ ($c>0$).

    $d$: the intrinsic decay rate of the predator, $Y$ (its decay rate in the absence of the prey, $X$; $d>0$).

This model was originally proposed to explain the periodic population fluctuations of predators and prey in ecosystems, but it is also applicable to chemical reaction systems that exhibit similar dynamics, such as networks of autocatalytic or enzymatic reactions. The most notable feature of this model is that, for an appropriate range of parameter values, the populations (concentrations) of $X$ and $Y$ oscillate periodically.

### A-1-2. Chemical Reaction Scheme for the Lotka-Volterra Equations
The standard Lotka-Volterra differential equations can be represented by the following elementary chemical reaction scheme.
    **1. Self-replication of Prey ($X$)** This reaction corresponds to the growth term, $aX$.

$$X \xrightarrow{a} 2X$$

This describes a molecule of prey $X$ replicating itself. In the context of the full model, the rate constant $a$ is determined by the concentration of a nutrient source, $[A]$, and a rate constant, $k_1$, such that $a = k_1[A]$.

**2. Predation and Replication of Predator ($Y$)** This single reaction corresponds to both the prey consumption term, $-bXY$, and the predator growth term, $+cXY$.

$$X + Y \xrightarrow{k_2} 2Y$$

This describes a molecule of predator $Y$ consuming a molecule of prey $X$ to produce two molecules of $Y$. The rate constant $k_2$ corresponds to both parameters $b$ and $c$ in the standard equations.

**3. Decay of Predator ($Y$)** This reaction corresponds to the decay term, $-dY$.

$$Y \xrightarrow{d} P$$

This describes a molecule of predator $Y$ decomposing into an inert product $P$. The rate constant $d$ is equivalent to $k_3$ in the full model.

Applying the law of mass action to this reaction scheme yields the standard Lotka-Volterra equations. The correspondence between the general parameters and the specific rate constants used in our model's implementation is as follows: $a = k_1[A]$, $b = k_2$, $c = k_2$, and $d = k_3$.

### A-1-3. The Oscillation Center

The center of oscillation for $X$ and $Y$ in the standard Lotka-Volterra equations is the system's **non-trivial fixed point** (or **equilibrium point**). This is the point where the rates of change are zero for both species simultaneously, i.e., where $dX/dt = 0$ and $dY/dt = 0$, for $X \neq 0$ and $Y \neq 0$.

From the system of equations:

$$X(a - bY) = 0$$

$$Y(cX - d) = 0$$

the non-trivial solution is found by solving:

$$a - bY = 0 \Rightarrow Y_s = \frac{a}{b}$$

$$cX - d = 0 \Rightarrow X_s = \frac{d}{c}$$

Therefore, the center of oscillation, which is the non-trivial fixed point $(X_s, Y_s)$, is given by:

$$(X_s, Y_s) = \left(\frac{d}{c}, \frac{a}{b}\right)$$

Substituting our model's specific parameters ($a = k_1[A]$, $b = k_2$, $c = k_2$, $d = k_3$):

$$X_s = \frac{k_3}{k_2}$$

$$Y_s = \frac{k_1[A]}{k_2}$$

This relationship is crucial as it demonstrates that the equilibrium concentration of Y, denoted as $Y_s$, can be controlled by the concentration of the nutrient source, [A], and the rate constant $k_1$. Since the production rate of the output substance P is given by $k_3[Y]$, its average production rate (i.e., the center of its oscillation) can therefore be actively controlled by modulating [A] or $k_1$.

This type of fixed point is known as a **center**. The trajectories around it are neutrally stable, meaning they neither spiral toward nor are repelled from the fixed point. Instead, the system oscillates periodically along a closed orbit whose specific shape and size are determined by the initial conditions.

### A-1-4. Numerical Method

This model uses the **4th-order Runge-Kutta method (RK4)**. This method was chosen over simpler techniques, such as the Euler method, as it provides higher accuracy and stability, even with a relatively large time step (*dt*).

### Appendix A-2. An Early Algorithmic Approach (Heuristic Model)

*Note: This section describes an early, heuristic algorithm that was considered during the model's development. The final simulations presented in this paper are based on the more chemically plausible "Antagonistic Memory Molecule Model," which is detailed in Appendix A-3.*

In the Lotka-Volterra oscillation model, an external nutrient, A, drives the chemical oscillation, causing the concentrations of substances X and Y to vary periodically. Consequently, the production rate of the product, P, also oscillates. This production rate is linked to the cell's pigmentation, which in turn affects its temperature.

The conceptual goal of this model is to enable a cell to spontaneously approach its optimal temperature, $T_{opt}$. This is achieved by having the cell "explore" its environment via the oscillations in the production rate of P. An adjustment loop then modifies the cell's albedo and the LV parameter $k_1$ based on the outcome of this exploration. This entire process operates under the key premise that the cell has no explicit knowledge of its own temperature ($T_{cell}$) or the optimal temperature ($T_{opt}$).

In this heuristic approach, a mechanism varies the albedo in proportion to the oscillating production rate of P. Instead of causing large swings, the albedo is designed to oscillate within a narrow range, allowing for fine-tuning of the cell temperature (e.g., within 1–2 °C). Specifically, the cell temperature would be adjusted according to the following logic, evaluated over one oscillation cycle of P:
- **If trends are anti-correlated** (i.e., the trend of P is positive while the trend of the

replication rate, $\beta$, is negative, or vice versa):
- Increase $k_1$ and decrease albedo (action to raise cell temperature).
- **If trends are correlated** (i.e., the trends of $P$ and $\beta$ are both positive or both negative):
  - Decrease $k_1$ and increase albedo (action to lower cell temperature).

*(Note: Since a higher production rate of P is linked to a higher albedo, this change in albedo results in an opposite change in cell temperature.)*

Through this feedback mechanism, the cell's temperature would asymptotically approach the optimal temperature by continuously shifting the oscillation center of $P$ in the direction that maximizes its replication rate. (This algorithm assumes a unimodal fitness landscape with a single optimal temperature. If multiple local maxima existed, a mechanism to introduce stochasticity would be necessary to escape them; however, this feature was not included in this conceptual model.)

This hypothetical $k_1$ adjustment loop would, therefore, realize a process of observation → belief update → action optimization.

## Appendix A-3. The Antagonistic Memory Molecule Model: A Chemically Plausible Implementation of Inference

The heuristic algorithm detailed in Appendix A-2, which relies on conditional "if-then" logic, was an initial consideration . Although this approach could theoretically guide the system to an optimal temperature, it has two significant drawbacks: it is algorithmically complex and, more importantly, lacks chemical plausibility .

To address these issues and to enhance both chemical feasibility and implementation simplicity, we developed the "Antagonistic Memory Molecule Model," which was used for all simulations presented in this study . The guiding principle behind this new model was to replace the discrete, algorithmic decision-making process with a continuous, dynamic system governed by simple chemical reactions. Specifically, the mechanism for adjusting the parameter $k_1$ and the baseline albedo—a process analogous to Bayesian inference—was reformulated entirely in terms of chemical reaction equations .

### A-3-1. The Antagonistic Memory Molecule Model

This model introduces two types of hypothetical "memory molecules," whose production and degradation are tied to the cell's overall success. The balance in the concentrations of these molecules continuously adjusts the parameter $k_1$ and the baseline albedo value (albedo center).

- $M_{good}$: A molecule that accumulates evidence that the cell's state is trending in a "good" direction (i.e., the self-replication rate, $\beta$, is improving).
- $M_{bad}$: A molecule that accumulates evidence that the cell's state is trending in a "bad" direction (i.e., the self-replication rate, $\beta$, is worsening).

The model's logic can be interpreted as the following set of processes:

**1. Production of Memory Molecules** The production of $M_{good}$ and $M_{bad}$ is controlled by the temporal correlation between the trend in the production rate of $P$ ($P$ rate) and the trend in the self-replication rate, $\beta$.

- When $P$ rate and $\beta$ are **positively correlated** (i.e., both are increasing or both are decreasing), the production of $M_{good}$ is promoted.
- Conversely, when they are **anti-correlated** (i.e., one is increasing while the other is

decreasing), the production of $M_{bad}$ is promoted.

**2. Degradation of Memory Molecules** Both $M_{good}$ and $M_{good}$ are assumed to degrade naturally at a constant rate. This ensures that the memory is not permanent, allowing the system to gradually forget past information.

$$M_{good} \rightarrow \text{degradation products}$$
$$M_{bad} \rightarrow \text{degradation products}$$

**3. Adjustment of $k_1$ and Baseline Albedo** The parameter $k_1$ and the baseline albedo (albedo center) are continuously adjusted based on the concentration difference between $M_{good}$ and $M_{bad}$.
- If $[M_{good}] > [M_{bad}]$: $k_1$ is decreased and the baseline albedo is increased (action to lower cell temperature).
- If $[M_{good}] < [M_{bad}]$: $k_1$ is increased and the baseline albedo is decreased (action to raise cell temperature).

The advantages of this new model include the following:
1. **Chemical Plausibility:** This model can be interpreted in terms of mechanisms that are ubiquitous in biochemistry, such as enzymatic activation/inhibition and feedback control loops.
2. **Simplicity and Robustness:** It eliminates the need for complex, discrete logic, such as the explicit detection of oscillation cycles or conditional "if-then" statements. Instead, the concentrations of the two memory molecules change continuously, and their balance smoothly adjusts the system's parameters. This dynamic process makes the system more robust to noise and leads to more stable operation.
3. **Continuous Learning:** The system does not need to wait for a full oscillation cycle to complete an update. It continuously evaluates the correlation between its internal state and success rate at each time step, allowing for more rapid and continuous adaptation.

Therefore, the Antagonistic Memory Molecule model represents the concept of a "Bayesian-like mechanism via simple chemical reactions" in a more direct and feasible manner.

### A-3-2. A Hypothetical Chemical Reaction Scheme for Correlation Detection

The correlation detection mechanism, which was represented abstractly by the differential equations in Section 4.3, can be implemented by a plausible series of elementary reactions. Here, we outline a hypothetical reaction scheme for this mechanism, using the production of $M_{good}$ as an example. Although our simulation employed the discrete algorithm described in **Appendix A-3-1**, this scheme demonstrates how its core function—detecting the correlation between two trends—can emerge from standard chemical kinetics.

**Step 1: Conversion of Trends into Chemical Signals** It is assumed that the rate of change of the internal state ($x$) and the rate of change of the success rate ($y$) drive the production of signal molecules, $S_x$ and $S_y$, respectively.
- Production of signal molecules (driven by positive trends):

$$\emptyset_X \xrightarrow{x} S_x$$
$$\emptyset_Y \xrightarrow{y} S_y$$

- Degradation of signal molecules:

$$S_x \xrightarrow{k_{d1}} \emptyset$$
$$S_y \xrightarrow{k_{d2}} \emptyset$$

*(Note: For simplicity, only the case where trends are positive is described. A similar system, in which other signal molecules such as Sx' and Sy' are produced when x and y are negative, can be readily conceived.)*

**Step 2: Cooperative Signal Detection by an Enzyme (An "AND" Gate)** The enzyme responsible for producing $M_{good}$ exists in an inactive state, $E_{inactive}$. It transitions to its active form, $E_{active}$, only when both signal molecules, $Sx$ and $Sy$, bind to it cooperatively.

$E_{inactive} + Sx \rightleftharpoons E_{Sx}$ (Reversible binding)
$E_{inactive} + Sy \rightleftharpoons E_{Sy}$ (Reversible binding)
$E_{Sx} + Sy \rightleftharpoons E_{active}$ (Cooperative binding)
$E_{Sy} + Sx \rightleftharpoons E_{active}$ (Cooperative binding)

**Step 3: Production of the Memory Molecule** Only the active form of the enzyme, $E_{active}$, possesses the catalytic function to convert a generic *Substrate* into $M_{good}$.

$$E_{active} + \text{Substrate} \xrightarrow{k_{cat}} E_{active} + M_{good}$$

This series of reactions demonstrates that $M_{good}$ is produced efficiently only when $Sx$ and $Sy$ are present simultaneously. It thus represents one plausible example of how the multiplicative $x \times y$ term in the abstract differential equation can be implemented through concrete chemical kinetics.

**Appendix A-4. Model Implementation Details**
**A-4-1. Implementation Overview**
- **Technology:** The simulation was implemented using HTML and JavaScript.
- **Model Dimensionality:** The model is one-dimensional.
- **Environment:** The external sunlight intensity is set to fluctuate exogenously, following a sinusoidal pattern.
- **Population Dynamics:** The model simulates the total cell population, which starts from a fixed initial value. However, the absolute population size is treated as a supplementary quantity. The primary focus of this study is the homeostatic regulation of the cell's replication rate, not complex population dynamics. To stabilize the population for analysis and prevent unrealistic exponential growth or extinction, cell growth is constrained by a logistic function.

**A-4-2. Parameter List**
The standard parameters used in the simulations are listed below.

| Category | Parameter | Value |
|---|---|---|
| Environment | Sunlight Intensity, L (Average) | 0.7 |
| | Sunlight Intensity, L (Amplitude) | 0.2 |
| | Sunlight Intensity, L (Period) | 200 |
| Cell | Initial Population, N0 | 50 |
| | Carrying Capacity, K | 1000 |
| | Lifespan | 5 |
| | Optimal Replication Temp., Topt | 22.0 °C |
| | Replication Temp. Range (Half-width) | 15 |
| Lotka-Volterra (LV) Oscillator | Nutrient Concentration, [A] | 1 |
| | Initial k1 | 1 |
| | k2 | 0.1 |
| | k3 | 0.4 |
| | Initial X0 | 10 |
| | Initial Y0 | 5 |
| Adaptation & Conversion | k1 Adjustment Coefficient | 0.035 |
| | Minimum k1 | 0.1 |
| | Maximum k1 | 2.5 |
| | Albedo Adjustment Coefficient | 0.035 |
| | Minimum Albedo (Black) | 0.05 |
| | Maximum Albedo (White) | 0.95 |
| | P-rate to Albedo Oscillation Coeff. | 0.003 |
| | Heat Exchange Coefficient, q | 20 |
| | Moving Average Window for β | 20 |
| | Memory Molecule Production Rate | 0.2 |
| | Memory Molecule Degradation Rate | 0.05 |
| Simulation Control | Total Time, T | 1000 |
| | Time Step, dt | 0.05 |

### A-4-3. Source Code Availability

The complete source code for the simulation, which is based on the Antagonistic Memory Molecule Model (Appendix A-3) and includes a graphical user interface, is available at the following GitHub repository:
https://github.com/Takeshi-Ishida/Emergence-of-Active-Inference-from-a-Chemical-Oscillator

The main simulation can be run by opening the following file in a web browser:

**Appendix A-5. Supplementary Results**

**A-5-1. Verification of Model Operation**

This section verifies that the model operates according to the specified algorithm by examining the dynamics of a representative simulation run, as shown in **Supplementary Figure 1**. The left panel of the figure shows the results for the first 200 time steps, while the right panel provides a magnified view of steps 100–120.

**Description of Left Panel (Overall Behavior)**

- **A. Lotka-Volterra Oscillation:** The internal LV model generates stable oscillations in the concentrations of $X$ and $Y$, which in turn drives the oscillation in the production rate of $P$.

- **B. Temperature Convergence:** The cell temperature progressively converges toward the set optimal temperature (22°C), effectively decoupling from the fluctuating external water temperature.

- **C. Replication Rate:** Consequently, the replication rate, $\beta$, is maintained near its maximum value of approximately 1.0.

- **D. Memory Molecules:** The concentrations of the memory molecules, $M_{good}$ and $M_{bad}$, fluctuate antagonistically to regulate the system and maintain the optimal temperature.

- **Initial Phase:** The period up to time step 70, where the cell temperature deviates significantly from the optimum and $\beta$ is low, is considered an initial adjustment phase due to the low initial values of the memory molecules.

**Description of Right Panel (Core Adaptive Logic)** The magnified view details the core adaptive mechanism at work:

- The **yellow-shaded region** highlights a phase where the production rate of $P$ and the replication rate $\beta$ are positively correlated (both increasing). As per the model's rules, this promotes the production of $M_{good}$, which increases the albedo and in turn lowers the cell temperature toward the optimum.

- The **green-shaded region** highlights a phase where the trends are anti-correlated ($P$ production is decreasing while $\beta$ is increasing). This promotes the production of $M_{bad}$, which decreases the albedo and thereby raises the cell temperature, again moving it toward the optimum.

This step-by-step analysis confirms that the model is operating as designed, successfully adjusting its temperature via the correlation-based feedback loop.

A. Lotka-Volterra model chemical oscillation

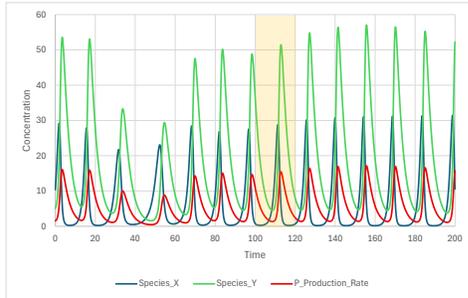
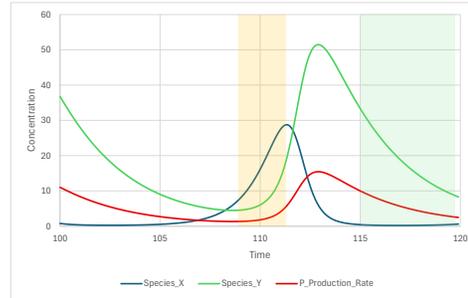

B. Transition of water and cell temperatures

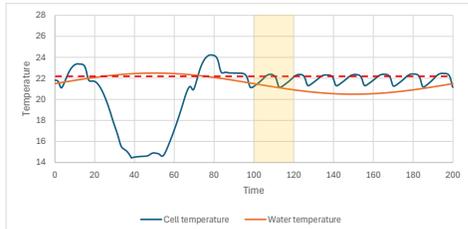
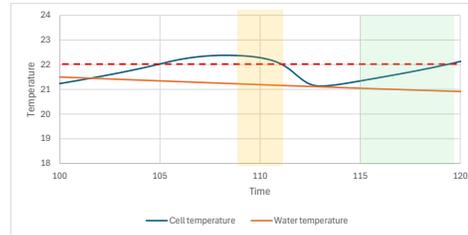

C. Transition of replication rate β and other parameters

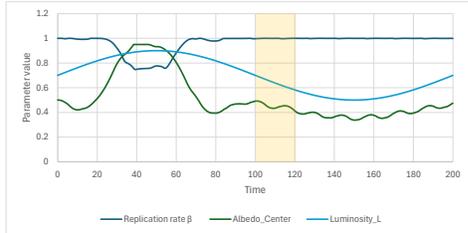
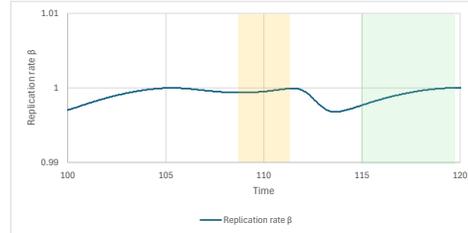

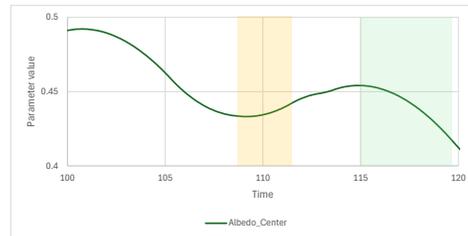

D. Transition of memory molecule counts

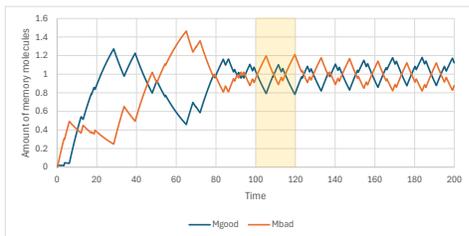
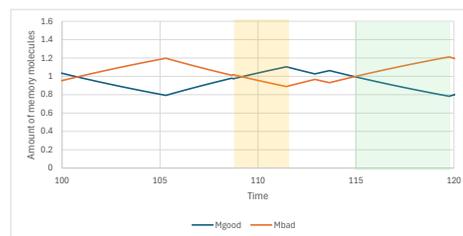

Supplementary Figure 1: Confirmation of the Algorithm

## A-5-2. Parameter Sensitivity Analysis

To assess the model's robustness, we performed a sensitivity analysis by varying several key parameters. The results, showing the mean (accuracy) and standard deviation (stability) of the steady-state cell temperature, are summarized in **Supplementary Table 1**.

**Supplementary Table 1. Effects of Various Parameters**

A. Heat Exchange Coefficient ($q$)

| Heat Exchange Coefficient | Optimal Temperature | Average cell temperature | Standard deviation of cell temperature |
|---|---|---|---|
| 10 | 22.00 | 21.98 | 0.304 |
| 20 | 22.00 | 22.06 | 0.597 |
| 30 | 22.00 | 22.17 | 0.868 |

B. Memory Molecule Production Rate

| Memory molecule generation rate | Optimal Temperature | Average cell temperature | Standard deviation of cell temperature |
|---|---|---|---|
| 0.05 | 22.00 | 21.93 | 0.485 |
| 0.1 | 22.00 | 22.06 | 0.597 |
| 0.15 | 22.00 | 23.30 | 1.339 |
| 0.2 | 22.00 | 22.16 | 0.893 |

C. Memory Molecule Degradation Rate

| Memory molecule decomposition rate | Optimal Temperature | Average cell temperature | Standard deviation of cell temperature |
|---|---|---|---|
| 0.025 | 22.00 | 22.79 | 1.085 |
| 0.050 | 22.00 | 22.06 | 0.597 |
| 0.075 | 22.00 | 22.08 | 0.536 |
| 0.100 | 22.00 | 21.98 | 0.520 |

D. P-rate to Albedo Conversion Coefficient

| Conversion factor from P production rate to albedo | Optimal Temperature | Average cell temperature | Standard deviation of cell temperature |
|---|---|---|---|
| 0.003 | 22.00 | 21.52 | 4.196 |
| 0.004 | 22.00 | 22.00 | 0.542 |
| 0.005 | 22.00 | 22.06 | 0.597 |
| 0.006 | 22.00 | 21.95 | 0.709 |
| 0.007 | 22.00 | 22.01 | 0.759 |

**Appendix A-6. Mapping the Model to the Free Energy Principle (FEP)**

**Supplementary Table 2. Mapping Key Model Components to FEP Concepts**

| FEP Concept | Corresponding Model Element | Description |
|---|---|---|
| Generative Model | The entire coupled system, comprising the LV oscillator and the learning mechanism. | A system that internally models the causal relationships between its internal state (LV oscillation), the action it generates (albedo change), and the resulting sensory input (replication rate, β). |
| Prior Beliefs | The implicit premise that a high replication rate (β) is desirable. | The agent does not explicitly know the target temperature, Topt. However, its learning mechanism is inherently designed to seek and maintain a high β, which corresponds to life's most fundamental prior: a preference for states conducive to survival. |
| Action | The dynamic change in the cell's albedo (pigmentation). | The physical modification of the phenotype in response to internal chemical oscillations, which actively alters the cell's thermal interaction with its environment. |
| Sensory States | The self-replication rate, β. | The sole feedback signal the system receives from the environment. It is not a direct physical quantity, such as temperature or luminosity, but rather highly low-dimensional information representing the system's overall 'success'. |
| Prediction Error | The concentration imbalance between the memory molecules, Mgood and Mbad. | Although there is no explicit prediction value, a decrease in the replication rate (β) following an action constitutes an 'unexpectedly poor outcome,' which is analogous to a prediction error. The accumulation of Mbad chemically represents a 'negative prediction error,' while the accumulation of Mgood represents a 'positive' or 'expected' outcome. The difference between their concentrations functions as the error signal for updating the internal model. |

**Appendix A-7. A Hypothetical Implementation via Allosteric Enzyme Control**

**Proposed Chemical Scenario**

The following is a hypothetical, yet plausible, chemical scenario for implementing the correlation-detection mechanism described in Section 4.3.

1. First, we assume that the rate of change of the internal state ($x$) and the rate of change of success ($y$) are represented by the concentrations of distinct chemical signals, let us call them $Sx$ and $Sy$. For example, in a system where the production rate of $Sx$ is proportional to $x$ and its degradation rate is constant, the concentration [$Sx$] would effectively track the value of $x$.

2. Next, we consider an enzyme, $E_{good}$, that produces the memory molecule $M_{good}$. This enzyme is proposed to have allosteric binding sites for both $Sx$ and $Sy$ (as depicted in Supplementary Figure 2) .

3. The crucial point is that $E_{good}$ only exhibits catalytic activity when $S_x$ and $S_y$ bind cooperatively. This mechanism functions as a chemical "AND" gate, activating the enzyme to produce $M_{good}$ only when the trends of the two signals align (i.e., when $x$ and $y$ are both positive).

4. Conversely, the production of $M_{bad}$ can be explained in a similar manner, by postulating a different enzyme, $E_{bad}$, that is activated only when the trends of $x$ and $y$ are opposed .

In this way, the sigmoid-like switching response, $\sigma(cxy)$, described in Section 4.3 can emerge from the fundamental biochemical principle of enzyme cooperativity. This suggests that the adaptive mechanism of our model is implementable as a simple chemical reaction network that obeys the law of mass action. This interpretation is theoretically supported by recent studies that demonstrate the mathematical equivalence between chemical kinetics and Bayesian inference (19).

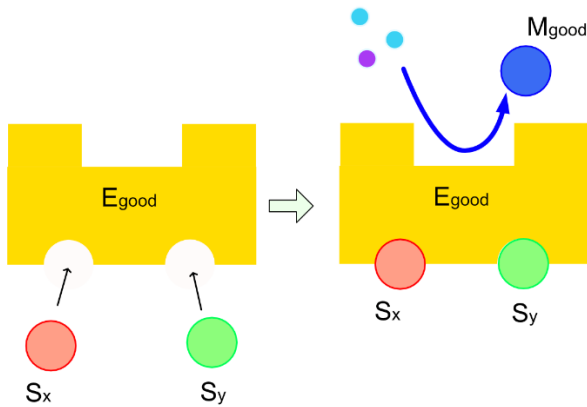

**Supplementary Figure 2.** Conceptual diagram of an allosteric enzyme for correlation detection. The enzyme $E_{good}$ has distinct binding sites for a signal $S_x$ (representing the trend in the internal state) and a signal $S_y$ (representing the trend in the success rate). The enzyme is activated to produce $M_{good}$ only upon the cooperative binding of both signals. This system functions as a chemical "AND" gate, generating $M_{good}$ only when the directions of the two trends align.